\documentclass[twocolumn,showpacs,superscriptaddress,amsmath,amssymb]{revtex4}
\usepackage{graphicx}

\let\c\undefined 
\let\k\undefined 

\newcommand{\fig}[1]{Fig.~\ref{#1}}

\newcommand{\eq}[1]{(\ref{#1})}

\newcommand{\Eq}[1]{Equation~(\ref{#1})}

\newcommand{\V}[1]{{\boldsymbol{#1}}}
\newcommand{\D}{\partial}

\newcommand{\f}{f_\mathrm{i}}
\newcommand{\w}{\overline\omega}
\newcommand{\e}{\overline\epsilon}
\newcommand{\E}{\overline{E}}
\newcommand{\F}{\overline{\V{\mathcal F}}}

\newcommand{\c}{k_1}
\newcommand{\k}{k_\mathrm{i}}
\newcommand{\K}{k_\mathrm{d}}
\newcommand{\G}{k_\mathrm{G}}
\newcommand{\Vk}{\V{k}_\mathrm{i}}

\newcommand{\dE}{\epsilon_\mathrm{dir}}
\newcommand{\iE}{\epsilon_\mathrm{inv}}
\newcommand{\dZ}{\eta_\mathrm{dir}}
\newcommand{\iZ}{\eta_\mathrm{inv}}
\newcommand{\ta}{\tau_{\alpha}}
\newcommand{\tE}{\tau_{\E}}
\newcommand{\tZ}{\tau_{Z'}}

\newcommand{\tw}{\tilde\omega}

\begin{document}

\title{Dynamics of Saturated Energy Condensation in Two-Dimensional Turbulence}

\author{Chi-kwan Chan}
\email{ckch@nordita.org}
\affiliation{NORDITA, Roslagstullsbacken~23, SE-10691 Stockholm, Sweden}

\author{Dhrubaditya Mitra}
\email{dhruba.mitra@gmail.com}
\affiliation{NORDITA, Roslagstullsbacken~23, SE-10691 Stockholm, Sweden}

\author{Axel Brandenburg}
\email{brandenb@nordita.org}
\affiliation{NORDITA, Roslagstullsbacken~23, SE-10691 Stockholm, Sweden}
\affiliation{Department of Astronomy, Stockholm University, 
  106~91 Stockholm, Sweden}

\date{\today}

\begin{abstract}
  In two-dimensional forced Navier--Stokes turbulence, energy cascades
  to the largest scales in the system to form a pair of coherent
  vortices known as the Bose condensate.
  We show, both numerically and analytically, that the energy
  condensation saturates and the system reaches a statistically
  stationary state.
  The time scale of saturation is inversely proportional to the
  viscosity and the saturation energy level is determined by both the
  viscosity and the force.
  We further show that, without sufficient resolution to resolve the
  small-scale enstrophy spectrum, numerical simulations can give a
  spurious result for the saturation energy level.
  We also find that the movement of the condensate is similar to the
  motion of an inertial particle with an effective drag force.
  Furthermore, we show that the profile of the saturated coherent
  vortices can be described by a Gaussian core with exponential wings.
\end{abstract}


\maketitle

\section{Introduction}

Two-dimensional (2D) hydrodynamic turbulence is fundamentally
different from its three-dimensional counterpart.
In 2D, small vortices can merge to form bigger coherent vortices.
This is because the equations of ideal hydrodynamics in two dimensions
have, in addition to energy, also enstrophy as a conserved quantity.
With an external force at an intermediate scale and viscous
dissipation, energy inversely cascades to larger length scales and
enstrophy directly cascades to smaller length
scales~\citep{1967PhFl...10.1417K, 1968PhFl...11..671L,
  1969PhFl...12..233B}.

\begin{figure}[b]
  \includegraphics[height=2.8in]{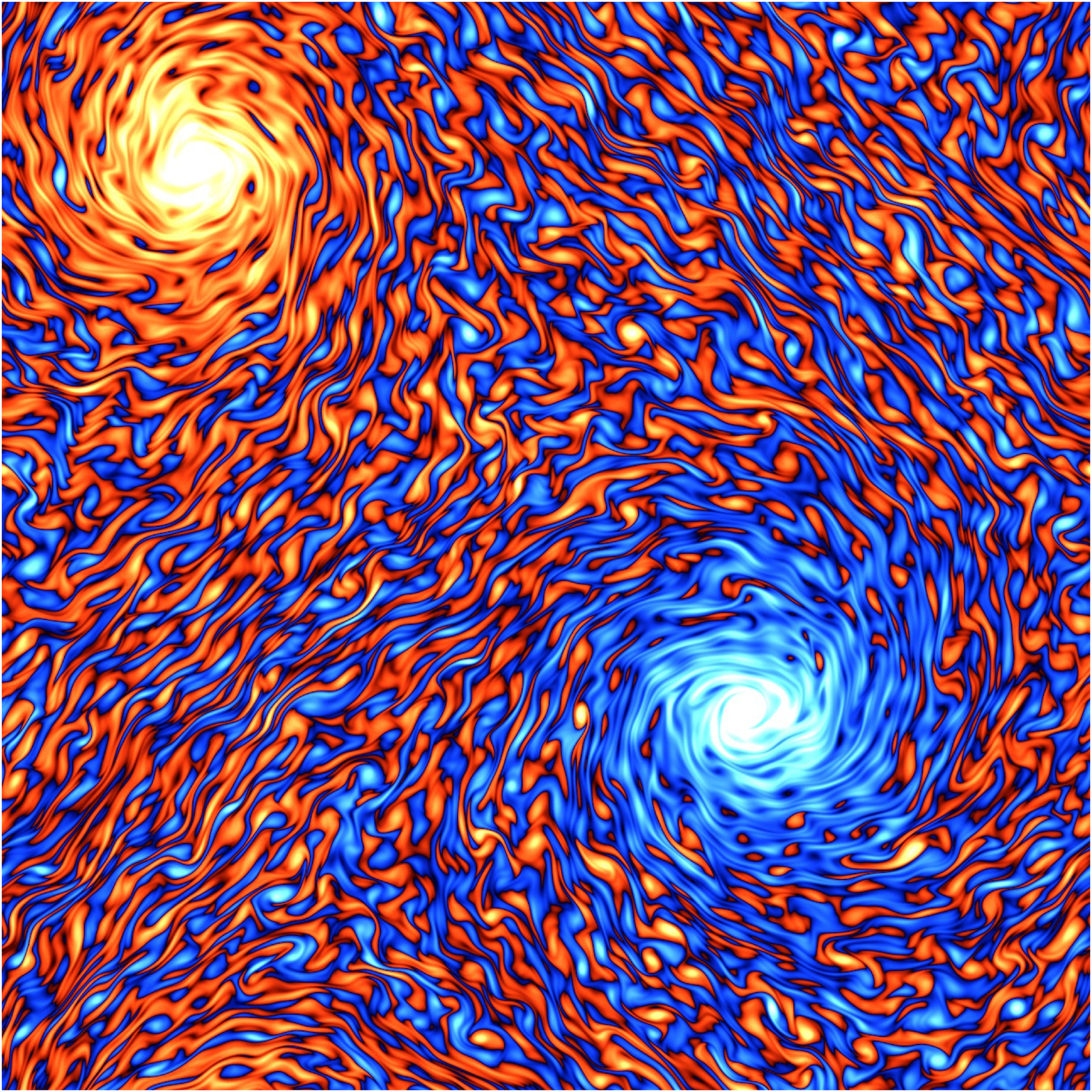}\hfill%
  \includegraphics[height=2.8in]{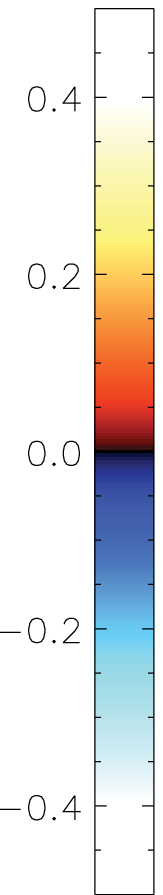}
  \caption{(Color online) Pseudocolor plot of vorticity showing Bose
    condensation.
    Red (the top-left vortex) and blue (the bottom-right vortex)
    represent positive and negative vorticity in physical space,
    respectively.
    The color scale is shown in the color bar on the right, which is
    chosen to make the fluctuation visible.
    The vorticity is normalized so that $\max|\omega| = 1$.}
  \label{fig:condex}
\end{figure}

Let us now consider 2D turbulence in a finite domain of size $L$.
The smallest wave number allowed in this system is $\c \equiv 2\pi/L$.
Due to the inverse cascade, energy piles up at~$k_1$ provided there is
no large-scale friction.
This phenomenon, sometimes called Bose condensation in 2D turbulence
(see \fig{fig:condex}), was first predicted by
\citet{1967PhFl...10.1417K}.
It was studied numerically by \citet{1983JPlPh..30..479H,
  1993PhRvL..71..352S, 1994JFM...274..115S, 2007PhRvL..99h4501C} and
experimentally by \citet{1998PhFl...10.3126P, 2008PhRvL.101s4504X}.

Following standard convention, we refer to the modes at $|\V{k}| = \c$
as the \emph{condensate} in this paper.
For a fixed non-zero viscosity, $\nu$, the energy of the condensate
vortices cannot grow without limit but saturate \citep{Eyink96,
  2004PhRvE..69c6303T}.
The saturation occurs at time scales of the order of $1/\nu\c^2$.
This is an unusual example of viscous effects playing an important
role in fluid turbulence at large length scales.
In this paper we show, by direct numerical simulations (DNS), that the
saturation of the condensate energy is not only determined by
viscosity but also by the forcing wave number $\k$.
Furthermore, we will demonstrate that the direct enstrophy cascade
must be well resolved for accurate numerical determination of the
saturation.

Motivated by the analogy between the formation of large-scale
structures in two-dimensional turbulence and the large-scale dynamo
process in three-dimensional helical magnetohydrodynamic turbulence
\citep{2001ApJ...550..824B}, we propose a simple three-scale model
which is able to capture the important aspects of our numerical
results.
We further show that the Lagrangian dynamics of the condensate
vortices can be described by Langevin equations for particles with
inertia.
Finally, we measure the profile of the saturated \emph{coherent}
vortices, which consist of the condensate and its higher harmonics.
The vorticity at the cores are well fitted by a sharp Gaussian; while
the wings fall off exponentially.

\begin{table*}
  \newcommand{\ml}{@{\ \ }l@{\ \ }}
\newcommand{\mc}{@{\ \ }c@{\ \ }}
\newcommand{\mr}{@{\ \ }r@{\ \ }}

\begin{tabular}{\ml\mr*{11}{\mc}}
  \hline\hline
  & \multicolumn{2}{c}{Inputs} & & \multicolumn{2}{l}{Predictions} & & \multicolumn{3}{c}{Results\ \ \ \ \ \ \ \ \ } & \\
  \!\!Name & $\nu$\ \ \ \ & $\k$ & $\K$ & $\tE$ & $E_\infty$ &Resolution\!\!\!&\ \ $\tilde{\tau}_{\overline{E}}$\ \ \ & $\tilde{E}_\infty$ & $\tilde{E}_\infty/E_\infty$ & Notes \\
  \hline
  \texttt{a16}               &        $10^{-3}$ & $ 16$ &  80 &  500 &  405 & $ 512^2$ &  483 &  323 & 0.80 & \\
  \texttt{a24}               &        $10^{-3}$ & $ 24$ &  91 &  500 &  356 & $ 512^2$ &  520 &  281 & 0.79 & \\
  \texttt{a32}               &        $10^{-3}$ & $ 32$ & 100 &  500 &  308 & $ 512^2$ &  582 &  234 & 0.76 & \\
  \hline
  \texttt{b16$^\mathtt{f}$}  & $5\times10^{-4}$ & $ 16$ & 113 & 1000 &  891 & $ 512^2$ &  --- &  --- &  --- &\!\!\!\!\!\!single-precision, crashed\!\!\!\!\!\!\\
  \texttt{b16}               & $5\times10^{-4}$ & $ 16$ & 113 & 1000 &  891 & $ 512^2$ &  922 &  696 & 0.78 & \\
  \texttt{b24}               & $5\times10^{-4}$ & $ 24$ & 129 & 1000 &  832 & $ 512^2$ &  947 &  643 & 0.77 & \\
  \texttt{b32}               & $5\times10^{-4}$ & $ 32$ & 142 & 1000 &  772 & $ 512^2$ &  970 &  580 & 0.75 & \\
  \hline
  \texttt{c16}               & $2\times10^{-4}$ & $ 16$ & 178 & 2500 & 2373 & $1024^2$ & 2223 & 1818 & 0.76 & \\
  \texttt{c32}               & $2\times10^{-4}$ & $ 32$ & 224 & 2500 & 2226 & $1024^2$ & 2287 & 1714 & 0.77 & \\
  \texttt{c64}               & $2\times10^{-4}$ & $ 64$ & 283 & 2500 & 1928 & $1024^2$ & 2409 & 1385 & 0.72 & \\
  \texttt{c64$^\mathtt{b}$}  & $2\times10^{-4}$ & $ 64$ & 283 & 2500 & 1928 &  $512^2$ & 1985 &  791 & 0.41 &\!\!\!\!\!\!converge to wrong answer\!\!\!\!\!\!\\
  \hline
  \texttt{d16}               &        $10^{-4}$ & $ 16$ & 252 & 5000 & 4859 & $1024^2$ & 4145 & 3581 & 0.74 & \\
  \texttt{d32}               &        $10^{-4}$ & $ 32$ & 317 & 5000 & 4691 & $1024^2$ & 4006 & 3349 & 0.72 & \\
  \hline
  \texttt{path$^\mathtt{i}$} & $5\times10^{-4}$ & $ 32$ & 142 & 1000 &  308 &  $512^2$ &  --- &  580 & 0.75 &\!\!\!\!\!\!restart from saturated \texttt{b32}\!\!\!\!\!\!\\
  \hline\hline
\end{tabular}

\let\ml\undefined
\let\mc\undefined
\let\mr\undefined

  \caption{List of simulations: The first 13 simulations are used to
    verify our three-scale model~\eq{sol:E1}.
    The last row describes 16 restart-runs which are initially in
    saturated states.
    They are used to verify the properties of $\Delta r^2$.
    The details of the simulations are described in
    Sec.~\ref{sec:sim}.}
  \label{tab:sim}
\end{table*}

\section{Numerical Simulations}
\label{sec:sim}

Let $\psi$ be the 2D (scalar) stream function.
The velocity is then $\V{u} = (\partial_y \psi, -\partial_x \psi)$ and
the $z$ component of the vorticity is $\omega = -\nabla^2\psi$.
We solve the 2D incompressible Navier--Stokes equations in the
vorticity--stream-function formulation, that is,
\begin{align}
  \D_t\omega - J(\psi, \omega) = \nu\nabla^2\omega + g, \label{eq:NS}
\end{align}
where the Jacobian determinant is given by
\begin{align}
  J(\psi, \omega) = (\D_x\psi)(\D_y\omega) - (\D_x\omega)(\D_y\psi).
\end{align}
We use periodic boundary conditions with $L = 2\pi$ so $\c = 1$, and
denote the Fourier transform of $\omega$ by $\omega_\V{k}$.

In Eq.~\eq{eq:NS}, $g$ is the $z$ component of the curl of an external
force.
Using $\langle\,\cdot\,\rangle$ to denote ensemble averages, the
Fourier transform of the force, $g_\V{k}$, is taken to be random and
white-in-time with zero means and variance
\begin{align}
  \big\langle g^*_\V{k}(s)\cdot g_\V{k}(t)\big\rangle =
  \f^2\k^2\;\delta(t - s).
\end{align}
It is achieved by choosing a random phase for $g_{\Vk}$ at each time
step.
In the above expression, $\f$ is the forcing amplitude and $\k$ is the
forcing wave number.
To ensure isotropy, we select $\V{k}$ randomly in a shell with radius
$\k$ and then round it off to the nearest grid point in Fourier space.
The effective width of the shell is approximately $\c$.
For this force, the average rates of energy and enstrophy inputs are
respectively $\f^2$ and $\f^2\k^2$.

We use a spectral Galerkin scheme in space and a low-storage
third-order Runge-Kutta/Crank-Nicolson time stepping scheme
\citep{1980JCoPh..35...48W}.
The time steps $\Delta t$ are chosen with a Courant number of 0.5.
The non-linear term is treated explicitly and the diffusion term is
treated implicitly.
The stochastic forcing is integrated by the Euler-Maruyama method
\citep{2001SIAMR..43..525H}.
We use $N \times N$ grid points with the Galerkin cutoff at $\G \equiv
\lfloor(N - 1) / 3\rfloor + 0.99$, where $\lfloor\,\cdot\,\rfloor$
denotes the floor function.
The value 0.99 is chosen such that the comparison $k^2 \le \G^2$ is
accurate enough even in single precision with $N \sim 2048$, although
almost all simulations in this paper are done with double precision.

Our code is implemented in \texttt{CUDA~C} and runs on graphics
processing units (GPUs), which are massively parallel ``stream''
processors.
With an \texttt{nVidia Tesla C2050} graphic card, our code is over an
order of magnitude faster than codes running on a single CPU core.
Because of communication overhead, our code outperforms
distributed-memory parallel codes running on 32 cores.
This speed-up allows us to integrate the problem over very long code
time and study the saturation of the condensate vortices.
We have released our Spectral Galerkin 2D code (\texttt{sg2}) as an
open-source project under the GNU General Public License version~3.
The project is hosted by the Google Code
\url{http://sg2.googlecode.com/}.

\section{Results}

We have performed a series of simulations as shown in
Table~\ref{tab:sim} with different sets of input parameters $\nu$ and
$\k$.
In all of them, the initial conditions are $\omega_\V{k}(t=0) = 0$.
We set the forcing amplitude at $\f = 1$ and choose the forcing wave
number $\k$ so that $\c \ll \k \ll \K$.
The dissipation wave number $\K$ is given by
\begin{align}
  \K \equiv \eta^{1/6}\nu^{-1/2}, \label{def:kd}
\end{align}
where $\eta$ is the forward enstrophy flux.
Further details of the runs are given in Table~\ref{tab:sim}.

\begin{figure}
  \includegraphics[width=\columnwidth,trim=16 8 8 16]{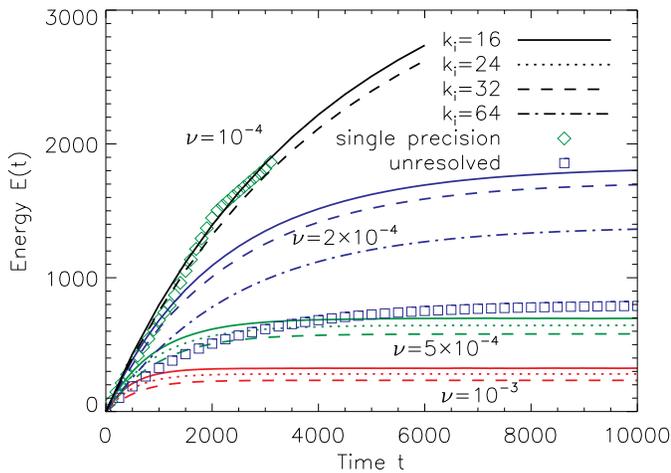}
  \caption{(Color online) Unnormalized energy evolution in simulations
    \texttt{a16} -- \texttt{d32}.
    The thick red (lowermost), green, blue, and black (uppermost)
    curves are for different viscosity $\nu$ (see labels); while
    solid, dashed, and dotted styles denote different forcing wave
    number $\k$ (see legend).
    The green diamonds and blue squares are for simulations
    \texttt{b16$^\mathtt{f}$} and \texttt{c64$^\mathtt{b}$}, respectively.
    They correspond to single precision and under-resolved
    simulations.}
  \label{fig:orgEt}
\end{figure}

In our simulations, energy inverse cascades to small Fourier modes
till $\c$ and forms a condensate at this mode.
We plot the total energy $E(t)$ for simulations \texttt{a16} --
\texttt{d32} in \fig{fig:orgEt}.
The different colors and line styles represent different choices of
viscosity $\nu$ (see labels) and forcing scale $\k$ (see legend).
The blue squares and green diamonds correspond to a single precision
(\texttt{b16$^{\mathtt{f}}$}) and an under-resolved
(\texttt{c64$^{\mathtt{b}}$}) simulations, which we will comment in
Sec.~\ref{sec:discussion}.
For intermediate time, the growth of energy of this condensate is
consistent with earlier results~\cite{2007PhRvL..99h4501C}.
Nevertheless, the asymptotic $E(t)$ becomes independent of time in all
the cases.
This saturation happens at a time scale determined by the viscous time
at the largest length scale (i.e., $\tau_\nu = 1/\nu\c^2$).
This requires very long integration times and hence has not been
explored by earlier simulations.
We are able to reach such late times by virtue of using a GPU code.
 
\begin{figure}
  \includegraphics[width=\columnwidth,trim=16 8 8 16]{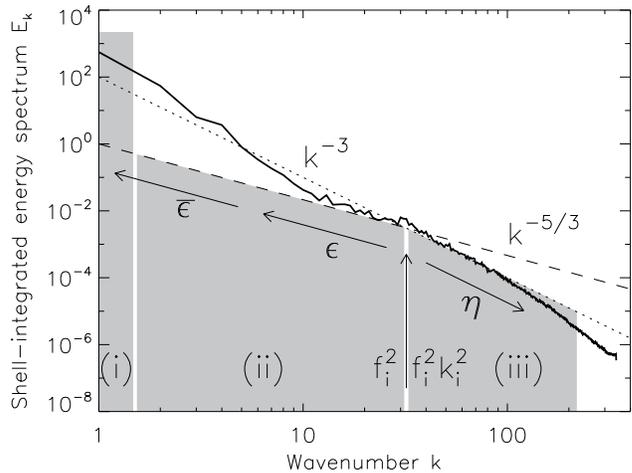}
  \caption{Energy spectrum in simulation \texttt{c32}.
    The thick solid curve is the energy spectrum $E_k$ computed at $t
    = 1000$ with bin size $\Delta k = \c$.
    The dashed line shows $k^{-5/3}$; while the dotted line is
    proportional to $k^{-3}$.
    The gray areas, which represent our three-scale model, are drawn
    in proper scales.
    The condensate at zone (i) contains a large amount of energy.
    The broken power law in zones (ii) and (iii) have slopes $-5/3$
    and $-3$, respectively.
    See Sec.~\ref{sec:evolution} for details of the model.}
  \label{fig:spec}
\end{figure}

At later times, most of the total energy comes from the energy of the
condensate.
Further understanding of the growth and saturation of the condensate
can be obtained by studying the energy spectrum.
We compute the shell-integrated energy spectrum $E_k$ in simulation
\texttt{c32} at $t = 1000$ using a fixed bin size $\Delta k = \c$.
The result is plotted as the thick solid curve in \fig{fig:spec}.
The condensate is well developed but not fully saturated.
The dashed and dotted lines in the figure are proportional to
$k^{-5/3}$ and $k^{-3}$, respectively.
Although the inertial ranges are narrow, the inverse energy cascade
and forward enstrophy cascade are consistent with spectral slope
$-5/3$ and $-3$.
However, the condensate strongly deviates the inverse cascade spectrum
at large scale.
This additional feature motivates us to introduce a simple three-scale
model.

\subsection{Energy and enstrophy evolution}
\label{sec:evolution}

Given the energy spectrum $E_k$, the total energy, enstrophy, and
palinstrophy can be computed by the one-dimensional integrals $E =
\int E_k dk$, $Z = \int k^2 E_k dk$, and $P = \int k^4 E_k dk$,
respectively.
Together with the average energy and enstrophy input, the evolution of
these quantities is governed by the following equations
\begin{align}
  \D_t E &= - 2 \nu Z + \f^2,     \label{eq:dE/dt} \\
  \D_t Z &= - 2 \nu P + \f^2\k^2. \label{eq:dZ/dt}
\end{align}

Our three-scale model is constructed in the following fashion.
We refer to the corresponding wave number ranges as zones (i) through
(iii) as shown in \fig{fig:spec}.
(i)~At the wave number of the box $\c$, we have the condensate modes,
which contain most of the energy at saturation.
We use the shorthand $\E(t)$ to denote the time-dependent energy at
this Fourier mode.
(ii)~In the range $\c < k \le \k$, the energy inverse cascades from
the forcing wave number $\k$ and forms an $E_k \sim k^{-5/3}$
spectrum.
Note that the enstrophy spectrum $Z_k = k^2 E_k$ peaks at the forcing
wave number $\k$.
(iii)~The forward enstrophy cascade in the small scales gives the
spectrum $E_k \sim k^{-3}$ with a sharp cutoff at $|\V{k}| = \K$.
It is necessary to assume this cutoff to prevent the ultraviolet
divergence of the total enstrophy and palinstrophy, although not total
energy.

Let us now denote by $\E$ the energy in the $|\V{k}| = \c$ mode (i.e.,
the condensate) and by $E'$ the energy of the rest of the system
(i.e., $E = \E + E'$).
Following standard mean-field theory, we call them, respectively, the
mean and fluctuating parts, even though $E'$ contributes to the
coherent structure as we will show in Sec.~\ref{sec:shape}.
Similar notations are applied to the enstrophy and palinstrophy too.
We can then split the energy equation~\eq{eq:dE/dt} into two parts,
namely,
\begin{align}
  \D_t \E &= -2\nu\c^2\E     + \e, \label{eq:dE1/dt}\\
  \D_t E' &= -2\nu Z' + \f^2 - \e, \label{eq:dE'/dt}
\end{align}
where $\e$ is the inverse energy transfer rate to the condensate.
Similarly, with the help of Eq.~\eq{eq:dE1/dt}, we can rewrite
Eq.~\eq{eq:dZ/dt} as
\begin{align}
  \D_t Z' = -2\nu P' + \f^2\k^2 - \c^2\e. \label{eq:dZ'/dt}
\end{align}
The last term $\c^2\e$ describes the enstrophy transfer rate to the
condensate.
Its existence implies that, although most of the enstrophy is
transferred toward smaller scales, there is a small \emph{leakage}
towards the largest scale.

Equations~\eq{eq:dE1/dt} through \eq{eq:dZ'/dt} are exact but not
closed.
Hence, it is not yet possible to solve them simultaneously.
We parametrize the fluctuating palinstrophy and enstrophy by $\gamma$
and $\Gamma$ such that
\begin{align} 
  P' \equiv P - \c^4\E &= \gamma\K^2 Z', \\
  Z' \equiv Z - \c^2\E &= \Gamma\k^2 E'.
\end{align}
Substituting the above equations into Eqs.~\eq{eq:dE'/dt} and
\eq{eq:dZ'/dt} we obtain two simultaneous equations.
We then eliminate $\e$ between them to obtain a dynamical equation for
$Z'$, which can be solved to obtain
\begin{align}
  Z'(t) = Z'_\infty \left[1 - e^{-(t - t_0)/\tZ}\right],
\end{align}
where we have used the initial condition $Z'(t_0) = 0$ and
\begin{align}
  Z'_\infty \equiv \frac{\f^2}{2\nu}
  \frac{\k^2 - \c^2}{\gamma\K^2 - \c^2}, \label{sol:Zinf}
\end{align}
with characteristic time scale
\begin{align}
  \tZ \equiv \frac{1}{2\nu\Gamma\k^2}
  \frac{\Gamma\k^2 - \c^2}{\gamma\K^2 - \c^2}.
\end{align}

Now note that the enstrophy saturation time scale $\tZ \approx 1 /
2\nu\gamma\K^2$ is substantially faster than the energy saturation
time scale, which is of the order of $\nu\c^2$.
Hence, for simplicity, we can replace $\e$ by its value at late time
in Eq.~\eq{eq:dE1/dt}, which can be obtained by replacing $Z'$ by
$Z'_{\infty}$ in Eq.~\eq{eq:dE'/dt},
\begin{align}
  {\e}_\infty \equiv \lim_{t\rightarrow\infty}\e(t)
  = \f^2 - 2\nu Z'_\infty = \f^2\frac{\gamma\K^2 - \k^2}{\gamma\K^2 - \c^2}.
\end{align}
Substituting this back into Eq.~\eq{eq:dE1/dt} and solving for $\E$,
we obtain
\begin{align}
  \E(t) = {\E}_\infty\Big[1 - e^{-(t - t_0)/\tE}\Big],  \label{sol:E1}
\end{align}
where we have used the initial condition $\E(t_0) = 0$.
This solution saturates to
\begin{align}
  {\E}_\infty \equiv \frac{\f^2}{2\nu \c^2}
  \frac{\gamma \K^2 - \k^2}{\gamma \K^2 - \c^2} \label{sol:Einf}
\end{align}
with a characteristic time scale
\begin{align}
  \tE \equiv 1/2\nu \c^2.
\end{align}
In the saturation level~\eq{sol:Einf}, the first part $\f^2/2\nu\c^2$
can be easily derived by dimensional arguments; while the correction
factor $(\gamma\K^2 - \k^2)/(\gamma\K^2 - \c^2) < 1$ comes from our
three-scale model.
This result is consistent with the energy bounds derivation by
\citet{Eyink96}.

\begin{figure}
  \includegraphics[width=\columnwidth,trim=16 8 8 16]{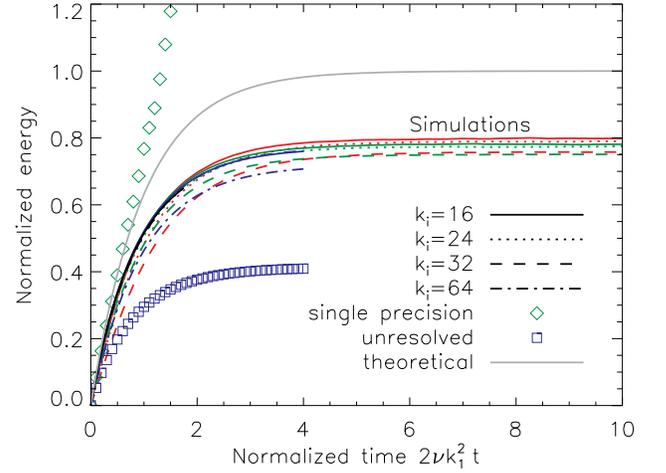}
  \caption{(Color online) Normalized energy evolution in simulations
    \texttt{a16} -- \texttt{d32}.
    The thick red, green, blue, and black curves (all under the
    ``Simulations'' label) are for different viscosity (see also
    labels in \fig{fig:orgEt}); while solid, dashed, and dotted styles
    denote different forcing wave number $\k$ (see legend).
    We normalize the energy and time using the results from our
    three-scale model.
    Almost all the curves collapse onto each other.
    The gray solid curve, green diamonds, and blue squares are the
    theoretical prediction, \texttt{b16$^\mathtt{f}$}, and
    \texttt{c64$^\mathtt{b}$}, respectively.}
  \label{fig:Et}
\end{figure}

To obtain the values of ${\E}_\infty$, we need to solve for $\K$ and
$\gamma$.
For the former, we combine definition~\eq{def:kd} and the forward
enstrophy flux in Eq.~\eq{eq:dZ'/dt} to obtain
\begin{align}
  \K = \f^{1/3}(\k^2 - \c^2)^{1/6}\nu^{-1/2}.
\end{align}
For the latter, we assume that in the inverse cascade $E_k \sim
k^{-5/3}$, and in the direct cascade $E_k \sim k^{-3}$.
These two power laws must match at $k = \k$.
Using these, we can compute the dimensionless number
\begin{align}
  \gamma \approx \frac{2}{3 + 4\ln(\K/\k)} \lesssim 1. \label{app:gamma}
\end{align}

With this, the model no longer has any free parameter.
In \fig{fig:Et}, we re-plot $E(t)$ with the vertical axis normalized
by ${\E}_\infty$ and also scale the time axis by $\tE$.
This collapses all the different time traces, thus providing support
to our three-scale model.
Note that the non-trivial correction factor given in Eq.~\eq{sol:Einf}
is necessary to obtain this collapse.
Note also that, the expressions for ${\E}_\infty$ and $Z'_\infty$ (but
not the time scales) can be derived directly by demanding that we
reach a stationary state at late times.
Imposing the conditions $\D_t E = \D_t Z = 0$ to Eqs.~\eq{eq:dE/dt}
and \eq{eq:dZ/dt}, Eqs.~\eq{sol:Einf} and \eq{sol:Zinf} follow
immediately.

Although all the results from different simulations collapse onto each
other, this collapsed curve lies systematically below the analytical
solution shown as light gray curve in \fig{fig:Et}.
To estimate this discrepancy, we fit the numerical curves by the form
of Eq.~\eq{sol:E1} to find numerical values of the \emph{total} energy
and the saturation time scale.
These values are compared with their theoretical prediction in
Table~\ref{tab:sim}.
The time scales agree to a good accuracy.
The ratios between the theoretical $E_\infty$ and the fitted
$\tilde{E}_\infty$ are listed in the tenth column of
Table~\ref{tab:sim}.
The average of these ratios (not including \texttt{b16$^\mathtt{f}$},
\texttt{c64$^\mathtt{b}$}, and \texttt{path$^\mathtt{i}$}) is $\zeta
\equiv \langle \tilde{E}_\infty / E_\infty \rangle \approx 0.76$.
We speculate that this systematic difference exists because we have
ignored the feedback of the condensate to the spectrum in our
three-scale model.
We will discuss this speculation with more details in
Sec.~\ref{sec:discussion}.

Comments on the remaining run in Table~\ref{tab:sim} are now in order.
The blue squares are for the \texttt{c64$^\mathtt{b}$} run.
Although they look perfectly reasonable, the resolution of the
simulation, $512^2$, which gives $\G \approx 170$, is too low to
resolve the actual enstrophy dissipation wave number $\K = 283$.
Because spectral Galerkin methods have very little numerical
dissipation, the forward cascaded enstrophy cannot be removed, which
artificially increases $Z'_\infty$ and decrease $\e_\infty$.
Hence, the steady state energy $E_\infty$ converges to an (incorrect)
lower value.
In the language of numerical analysis, this is an inconsistent
numerical solution.

The green diamonds in \fig{fig:Et} are for the
\texttt{b16$^\mathtt{f}$} run, which uses single-precision
floating-point numbers to represent $\omega_\V{k}$ instead of double
precision.
The numerical solution behaves properly at the beginning, but starts
to diverge from the true solution around $t = 1000 = \tE$.
This is not surprising because the relative round-off error of
single-precision numbers is about $\sim 10^{-7}$.
A random walk of round-off error in $\omega_\V{k}$ leads to a linear
grow of error in the energy.
Indeed, there are roughly $10^7$ steps by the time we reach $t = \tE$,
which gives the order of unit error seen in the figure.
To date, a significant fraction of scientific computing in GPUs are
done with single precision.
This turns out to be insufficient for our purpose.

\subsection{Movement of Condensate Vortices}

After studying the saturation level and time scale, we focus on the
movement of the condensate vortices at late times.
The last row in Table~\ref{tab:sim} represents 16 different
simulations \texttt{path$^\mathtt{85}$} --
\texttt{path$^\mathtt{100}$}.
The superscripts in the names have the following meaning.
For \texttt{path$^\mathtt{i}$}, we pick the \texttt{i}-th output file
of \texttt{b32} and restart the simulation each time with a different
realization of the random force.
In other words, for the $\mathtt{i}$-th run the random number
generators are started with seed \texttt{i}.
We then evolve the solutions for another $t = 10 \tE$.
Because \texttt{b32} has reached a saturated state long before the
85-th output, the energy levels remain almost constant in all these
simulations.
The result is an ensemble of 16 saturated solutions \citep{move}.

\begin{figure}
  \includegraphics[width=\columnwidth,trim=16 8 8 16]{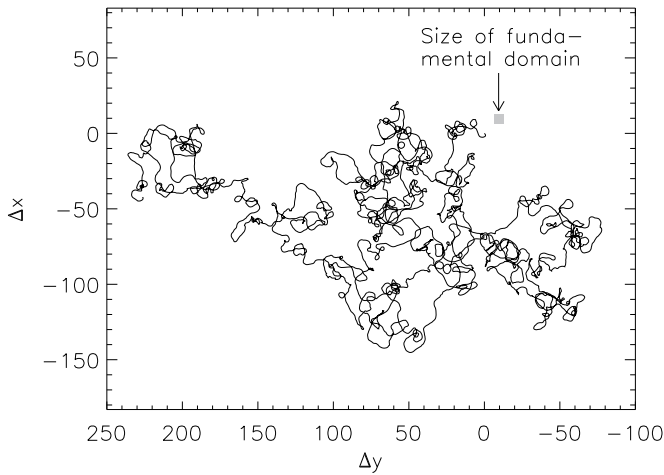}
  \caption{Unfolded trajectory of the positive condensate vortex in
    simulation \texttt{path$^\mathtt{100}$}.
    The computational domain is $[0,2\pi)^2$ (see gray box near the
    origin).
    We first unfold the trajectory into $\mathbb{R}^2$ and then shift
    the initial position to the origin.}
  \label{fig:upath}
\end{figure}

\begin{figure}
  \includegraphics[width=\columnwidth,trim=16 8 8 16]{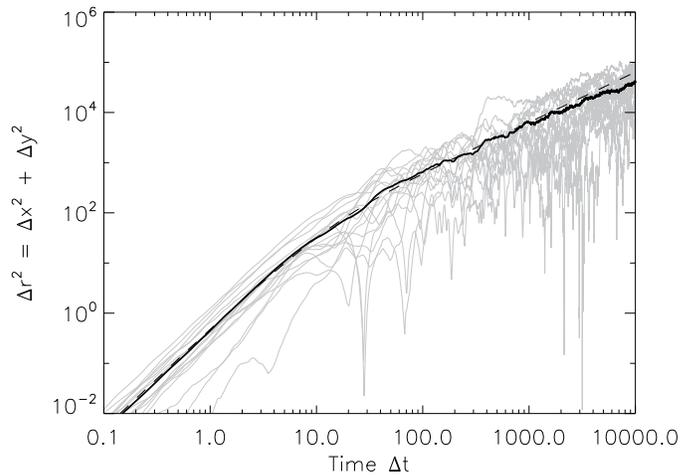}
  \caption{The gray curves are the displacement square $\Delta r^2$
    for all 16 \texttt{path$^\texttt{i}$} runs.
    The thick solid curve shows their average.
    The dashed curve, which is indistinguishable from the solid curve,
    is the solution~\eq{sol:r2} with parameter $\xi = 0.1$.}
  \label{fig:r2}
\end{figure}

The positions of the condensate vortices are simply given by the
phases of the $\c$ modes.
We unfold the trajectories from the computational domain $[0,2\pi)^2$
into $\mathbb{R}^2$ and shift their starting points to the origin.
We label respectively the $x$- and $y$-displacements, respectively, by
$\Delta x$ and $\Delta y$ and plot such a trajectory (for
\texttt{path$^\texttt{100}$}) in \fig{fig:upath}.
The small gray box near the origin shows the size of the original
domain $[0,2\pi)^2$.
With the unfolded trajectories, we compute the displacement square
$\Delta r^2 = \Delta x^2 + \Delta y^2$ for all 16
\texttt{path$^\texttt{i}$} runs.
For each simulation, we plot a gray curve in \fig{fig:r2}.
The thick solid curve shows their (ensemble) average.
In the following, we describe a way to model the motion of this
condensate.

We define the ``mean'' vorticity by
\begin{align}
  \w(\V{x}) \equiv \sum_{|\V{k}| = \c} \omega_\V{k} e^{i\V{k}\cdot\V{x}}
  \label{def:w}
\end{align}
so the fluctuating vorticity is $\omega' = \omega - \w$.
Averaging the Navier--Stokes equation, we obtain
\begin{align}
  \D_t\w + \V\nabla\cdot(\overline{\V u}\,\w) = 
  - \V\nabla\cdot\F + \nu\nabla^2\w.
\end{align}
In the above equation, $\F \equiv \overline{\V u\,\omega} -
\overline{\V u}\,\w$ is the space-dependent mean vorticity flux.
Note that the forcing $g$ is at small wave numbers $\k \gg \c$, so its
mean vanishes.
It is straightforward to show that the non-linear term
$\V\nabla\cdot(\overline{\V u}\,\w)$ vanishes identically.
Both the creation and the movement of the condensate, therefore, must
be due to the flux $\F$.

We now model this mean vorticity flux using the usual technique of
mean-field theory
\begin{align}
  \overline{\mathcal F}_i = \alpha_i\,\w + \beta_{ij}\D_j\w.
  \label{eq:flux}
\end{align}
The transport coefficients $\alpha_i$ is usually referred to as the
kinetic anisotropic alpha effect \citep{1987PhyD...28..382F,
  1989JFM...205..341S, 2001A&A...379.1153B}, and $\beta_{ij}$ is a
negative eddy diffusivity tensor.
Both $\alpha_i$ and $\beta_{ij}$ are constants in space because we
consider only the $\c$ modes.
The mean-field equation then becomes
\begin{align}
  (\partial_t + \alpha_i\D_i)\overline{\omega} =
  (\nu\delta_{ij}-\beta_{ij})\D_i\D_j\w. \label{eq:mf}
\end{align}
The coefficient $\alpha_i$ cannot change the amplitude of the
condensate.
It can be thought of as the \emph{Lagrangian velocity} of the pair of
condensate vortices.
The inverse energy cascade, therefore, must be described by
\emph{anti-diffusion}.

Expanding $\V\nabla\cdot\V\F$, there are only two independent Fourier
coefficients, $\overline{\mathcal F}_{x, \c\V{\hat x}}$ and
$\overline{\mathcal F}_{y, \c\V{\hat y}}$, that enter the mean-field
equation (the first subscript denotes component, the second one
denotes wave vector).
Comparing to our parametrization~\eq{eq:flux},
\begin{align}
  \alpha_x &= \mathrm{Re}
  \frac{\overline{\mathcal F}_{x, \c\V{\hat x}}}{  \omega_{\c\V{\hat x}}},&
  \alpha_y &= \mathrm{Re}
  \frac{\overline{\mathcal F}_{y, \c\V{\hat y}}}{  \omega_{\c\V{\hat y}}};
  \label{eq:a} \\
  \beta_{xx} &= \mathrm{Im}
  \frac{\overline{\mathcal F}_{x, \c\V{\hat x}}}{\c\omega_{\c\V{\hat x}}},&
  \beta_{yy} &= \mathrm{Im}
  \frac{\overline{\mathcal F}_{y, \c\V{\hat y}}}{\c\omega_{\c\V{\hat y}}}.
\end{align}
It is clear that $\langle\beta_{xx}\rangle = \langle\beta_{yy}\rangle
= \nu$ at saturation.

The value of $\V\alpha$, however, is much more difficult to obtain
because there is no constraint by conservation laws.
We can only perform a rough estimate: The flux $\F$ is a convolution
in Fourier space.
Given that $\c^2\E_\infty / Z'_\infty \gg 1$, the
condensate-fluctuation interaction dominates, so
\begin{align}
  \F \sim \V{u}_{\sqrt{2}\c}\w,
\end{align}
where $\V{u}_{\sqrt{2}\c}$ represents the typical value of a band-pass
filtered velocity.
Comparing the estimate with Eq.~\eq{eq:a} and employing the results
from our three-scale model, we obtain
\begin{align}
  \langle\alpha^2\rangle \sim \c E_{\sqrt{2}\c} \approx
  \frac{\f^2}{2\nu\K^2}\left(\frac{\k}{\c}\right)^{\!\!2/3} =
  \frac{\f^{4/3}}{2\c^{2/3}}. \label{est:a2}
\end{align}
Note that the same argument leads to $\langle\beta^2\rangle \sim
\c^{-2}\langle\alpha^2\rangle$.
Fortunately, it does not contradict $\langle\beta\rangle = \nu$.

Let us now take $\V{\alpha}$ to be statistically independent of $\w$.
\Eq{eq:mf} then becomes the equation of a passive scalar advected by a
random velocity $\V{\alpha}$.
As $\V{\alpha}$ has a stationary variance \eq{est:a2}, its simplest
model would be an Ornstein-Uhlenbeck process
\citep{1930PhRv...36..823U, 1943RvMP...15....1C, 1945RvMP...17..323W}
\begin{align}
  \D_t\V\alpha = - \V\alpha/\ta + \V\phi. \label{eq:particle1}
\end{align}
In the above phenomenological equation, $1/\ta$ is an effective drag
coefficient and $\V\phi$ is an effective stochastic forcing.
Because $\V\alpha$ is at a scale close to the condensate, we model the
effective forcing $\V\phi$ by Gaussian white noise with zero mean and
variance equal to $\xi\,\e$. Here $\xi$ is a tunable parameter and
$\e$ is the mean energy input to the condensate.
Standard It\=o calculus gives
\begin{align}
  \langle\V\alpha&(t)\rangle = \V\alpha(0) e^{-t/\ta}, \\
  \langle\V\alpha&(s)\cdot\V\alpha(t)\rangle = \alpha(0)^2 e^{-(t+s)/\ta}
  \nonumber\\
  &+ \xi\f^2 \frac{\gamma\K^2 - \k^2}{\gamma\K^2 - \c^2}\,\ta
  \left[e^{-(t-s)/\ta} - e^{-(t+s)/\ta}\right],
\end{align}
where the second equation is only valid for $s < t$.

By requiring $\langle\alpha(t)^2\rangle = \langle\alpha^2\rangle$ for
arbitrary large $t$, we can solve for the time scale
\begin{align}
  \ta = \frac{1}{2\xi\f^{2/3}\c^{2/3}}
  \frac{\gamma\K^2 - \c^2}{\gamma\K^2 - \k^2}.
\end{align}
Furthermore, starting with an expected initial condition $\alpha(0)^2
= \langle\alpha^2\rangle$ allows us to simplify the correlation to
\begin{align}
  \langle\V\alpha(s)\cdot\V\alpha(t)\rangle =
  \langle\alpha^2\rangle e^{-(t-s)/\ta}.
\end{align}
The displacement of the condensate can now be solved by using the
simple equation
\begin{align}
  \D_t\V r = \V\alpha. \label{eq:particle2}
\end{align}
The motion of an inertial particle in a fluid with Stokes drag is
described by the same pair of equations [Eqs.~\eq{eq:particle1} and
  \eq{eq:particle2}].
The role of effective velocity is played by $\V\alpha$ and the
effective Stokes time is given by $\ta$.

Let $\Delta\V r(t) \equiv \V r(t) - \V r(0)$ be the displacement from
the initial position, we have
\begin{align}
  \langle\Delta r(t)^2\rangle &= 2\int_0^t dt'
  \int_0^{t'}ds'\langle\V\alpha(s')\cdot\V\alpha(t')\rangle \nonumber\\
  &= 2\langle\alpha^2\rangle\ta \left[t + \ta(e^{-t/\ta} - 1)\right].
  \label{sol:r2}
\end{align}
Its asymptotic behavior is
\begin{align}
  \langle\Delta r(t)^2\rangle \approx \begin{cases}
    \;\;\langle\alpha^2\rangle\;t^2\;, & t < \ta; \\
    2\langle\alpha^2\rangle\ta t,      & t > \ta.
  \end{cases}
\end{align}
The early time behavior is completely determined by
$\langle\alpha^2\rangle$, while the free parameter $\xi$ controls only
the Brownian motion of the condensate at late time.

The dashed curve in \fig{fig:r2} is Eq.~\eq{sol:r2} with a parameter
$\xi = 0.1$, which corresponds to $\ta = 6.384$.
It is practically indistinguishable from the thick curve, except at
very late times ($\Delta t>3000$).
Note that at short times the condensate movement is independent of
$\xi$. Hence the agreement between the thick solid and dashed curves
for $\Delta t < \ta$ provides support to our phenomenological model.
Although we are not able to derive $\xi$, the late time agreement
suggests that the condensate motion is Brownian for $\Delta t > \ta$.

\subsection{Shape of Coherent Vortices}
\label{sec:shape}

\begin{figure}
  \includegraphics[width=\columnwidth,trim=16 8 8 16]{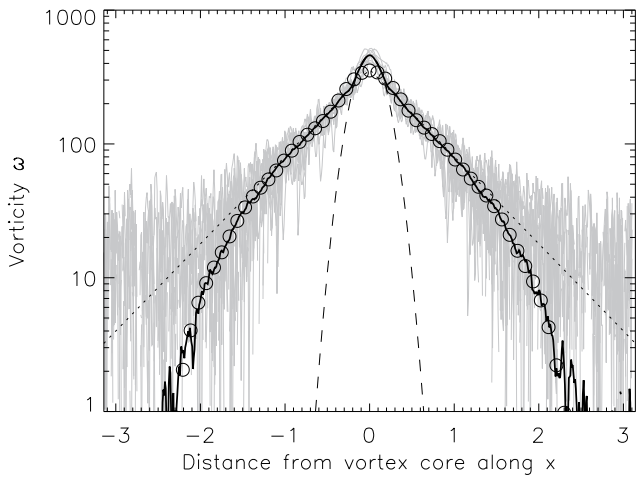}
  \caption{We apply coordinate transformations to shift the center of
    the positive condensate vortex to the origin in simulation
    \texttt{c32}.
    The noisy gray curves show the vorticity at several different
    snapshots along the $x$-axis (with $y = 0$).
    The black solid curve is an average over 1000 such profiles.
    It consists of a sharp Gaussian core (dashed curve) and
    exponential wings (dotted curve), although for $|x| \gtrsim \pi/2$
    it falls off super-exponentially.
    The open circles show the reconstruction from the Fourier
    coefficients listed in Table~\ref{tab:fit}.}
  \label{fig:shape}
\end{figure}

\begin{figure}
  \includegraphics[width=\columnwidth,trim=16 8 8 16]{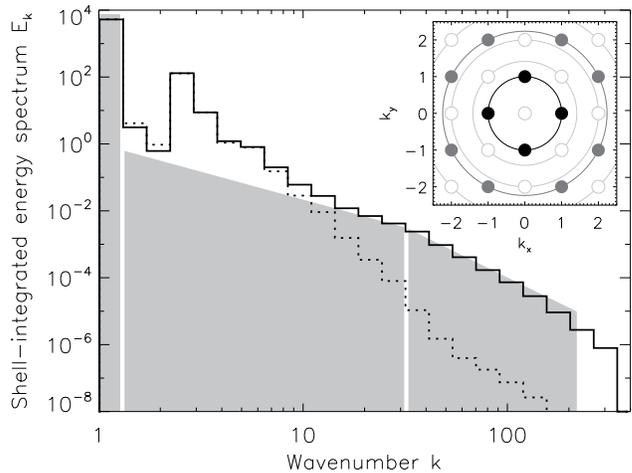}
  \caption{The solid and dotted histogram are the energy spectra $E_k$
    of the full ($\omega$) and the coherent ($\tw$) vorticity in
    simulation \texttt{c32} at $t = 10000$.
    The spectral bins have even width in log-scale.
    The valley and second peak right next to the condensate mode are
    physical.
    The valley corresponds to the $|\V{k}| = \sqrt{2}\c$ and $2\c$
    modes, which have even $k_x + k_y$ (open circles in inset).
    The second peak corresponds to the $|\V{k}| = \sqrt{5}\c$ modes,
    which are the first harmonics of the condensate (gray circles in
    inset).}
  \label{fig:harm}
\end{figure}

\begin{table*}
  \begin{tabular}{c|*{11}{r@{}l}}
  \hline\hline
      \multicolumn{1}{p{20pt}|}{\hspace*{8pt}$k_x$}
    & \multicolumn{1}{p{10pt}}{} & \multicolumn{1}{p{10pt}}{}
    & \multicolumn{1}{p{10pt}}{} & \multicolumn{1}{p{10pt}}{}
    & \multicolumn{1}{p{10pt}}{} & \multicolumn{1}{p{10pt}}{}
    & \multicolumn{1}{p{10pt}}{} & \multicolumn{1}{p{10pt}}{}
    & \multicolumn{1}{p{10pt}}{} & \multicolumn{1}{p{10pt}}{}
    & \multicolumn{1}{p{10pt}}{} & \multicolumn{1}{p{10pt}}{\hspace*{-8pt}$k_y$}
    & \multicolumn{1}{p{10pt}}{} & \multicolumn{1}{p{10pt}}{}
    & \multicolumn{1}{p{10pt}}{} & \multicolumn{1}{p{10pt}}{}
    & \multicolumn{1}{p{10pt}}{} & \multicolumn{1}{p{10pt}}{}
    & \multicolumn{1}{p{10pt}}{} & \multicolumn{1}{p{10pt}}{}
    & \multicolumn{1}{p{10pt}}{} & \multicolumn{1}{p{10pt}}{}\\
    & 0 & & 1 & & 2 & & 3 & & 4 & & 5 & & 6 & & 7 & & 8 & & 9 & & 10 & \\
  \hline
  0 &   &   & 28&.38&   &   &  5&.03&   &   &  1&.65&   &   &  0&.79&   &   &  0&.45&   &   \\
  1 &   &   &   &   & 10&.37&   &   &  2&.24&   &   &  1&.05&   &   &  0&.61&   &   &  0&.32\\
  2 &   &   &   &   &   &   &  2&.54&   &   &  1&.43&   &   &  0&.72&   &   &  0&.41&   &   \\
  3 &   &   &   &   &   &   &   &   &  1&.75&   &   &  0&.8 &   &   &  0&.52&   &   &  0&.14\\
  4 &   &   &   &   &   &   &   &   &   &   &  0&.85&   &   &  0&.6 &   &   &  0&.32&   &   \\
  5 &   &   &   &   &   &   &   &   &   &   &   &   &  0&.64&   &   &  0&.38&   &   &   &   \\
  6 &   &   &   &   &   &   &   &   &   &   &   &   &   &   &  0&.4 &   &   &   &   &   &   \\
  \hline\hline
\end{tabular}

  \caption{The ``upper triangular'' part of the Fourier coefficients
    of the coherent vortices, which are real because of Hermitian
    symmetry.
    The lower triangle and other quadrants are given by $\tw_{k_x,k_y}
    = \tw_{\pm k_y, \pm k_x}$ because of parity and discrete
    rotational symmetry.
    Coefficients that do not satisfy $|\tw_{\V{k}}| \gtrsim 10^{-3}
    |\tw_{\c}|$ are omitted in the table.
    We can reconstruct the shape of both the exponential wings and
    fast fall off in \fig{fig:shape} (see open circles in the figure)
    by using the shown coefficients.
    The sharp Gaussian core, however, requires higher wave numbers to
    be represented correctly.
    The omitted values exhibit an interesting pattern that can be
    derived by symmetry, see Sec.~\ref{sec:shape} for details.}
  \label{tab:fit}
\end{table*}

We can invert the procedure of computing the vortex trajectory to hold
fixed the positive condensate vortex at the origin of the
computational domain.
In \fig{fig:shape}, the noisy gray curves show ten different profiles
of the condensate along the $x$ axis (with $y = 0$).
Performing an average over 1000 such profiles, we obtain the black
solid curve.
To avoid confusion, we refer to this mean profile as the pair of
\emph{coherent vortices} and denote it by $\tw$, while the name
\emph{condensate vortices} is kept for~$\w$ [see
  definition~(\ref{def:w})].
The function $\tw(x, 0)$ can be well approximated by a sharp Gaussian
core (fitted by the dashed curve) and exponential wings (fitted by the
dotted curve) as shown in the figure.
There is no preference between positive and negative vorticity so
$\tw$ must change sign in the domain, which implies faster than
exponential fall off for $|x| \sim L/2$.
This non-trivial shape cannot be fully described by the 4 modes with
$|\V{k}|= \c$.

Nevertheless, it is possible to describe $\tw$ by a small number of
Fourier coefficients.
We find the Fourier coefficients of the coherent vortices satisfying
$|\tw_{\V{k}}| \gtrsim 10^{-3} |\tw_{\c}|$ and list the ``upper
triangular'' part in Table~\ref{tab:fit}.
The shown values are real because of Hermitian symmetry.
The lower triangle and other quadrants are given by $\tw_{k_x,k_y} =
\tw_{\pm k_y, \pm k_x}$ because of parity and discrete rotational
symmetry.
The open circles in \fig{fig:shape} show the reconstructed profile
from these coefficients.
Despite the fact that we only have a few modes, the approximation is
extremely good in both the exponential wings and the far tails where
the power falls off more rapidly.
The sharp Gaussian core, however, requires higher wave numbers to be
represented correctly because of its short length scale.

The omitted values in the table exhibit an interesting pattern.
The Fourier coefficients with even $k_x + k_y$ are significantly
smaller than the ones with odd $k_x + k_y$.
This is because the coherent vortices follow the same symmetry
properties of the condensate,
\begin{align}
  \underbrace{
    \tw(x, y) = \lefteqn{
      \overbrace{%
        \phantom{\tw(x + L, y + L) = -\tw(x + L/2, y + L/2).}%
      }^{\mbox{\scriptsize flip sign along diagonals}}
    }
    \tw(x + L, y + L)
  }_{\mbox{\scriptsize strictly periodic}}
  = -\tw(x + L/2, y + L/2).
  \raisetag{12pt}
\end{align}
Fourier transforming the above equation, we immediately obtain the
condition
\begin{align}
  \tw_{k_x, k_y} = - \tw_{k_x, k_y} (-1)^{k_x + k_y}.
\end{align}
Hence, the coherent vortices can only occupy modes with odd $k_x +
k_y$.

The above property has interesting implication for the energy spectrum
of 2D turbulence.
We graphically represent the Fourier amplitudes in the inset of
\fig{fig:harm}.
The most energetic $\c$ modes, which correspond to the value 28.38 in
Table~\ref{tab:fit}, are marked with filled circles.
Their ``higher harmonics'', with value 10.37, are in dark gray, which
form the second peak with radius $\sqrt{5}\c$ in the integrated
spectrum as shown by the solid histogram.
Other omitted modes with even $k_x + k_y$ are marked as open circles,
which form a spectral valley in the spectrum.

The spectral valley and the second peak next to the condensate in
\fig{fig:harm} are, to our knowledge, not seen in previous numerical
studies such as \citep{1993PhRvL..71..352S, 1994JFM...274..115S,
  2004PhRvE..69c6303T, 2007PhRvL..99h4501C}.
It is possible that these earlier studies have not reached the late
saturated stage we study, or perhaps because evenly spacing bins
``wash'' out the spectra as in \fig{fig:spec}.
In the later case, reducing the bin size to $k_1 / 2$ should make the
spectral valley and second peak visible, although the spectrum may
become noisier.
The spectrum obtained from atmospheric data \cite{1985RaSc...20.1339G}
or from direct numerical simulations with Ekman friction
\cite{2011PhRvL.106e4501P} would also not have these features as in
those cases energy from the large scales is removed by friction.
Nevertheless, we should remark that in Fig.~3 of
\citet{1983JPlPh..30..479H}, there is an indication of such a second
peak, although the authors did not comment on it.

As we show by the symmetry argument, the non-trivial shape of the
coherent vortices is responsible for these additional spectral
features.
To verify this, we plot the energy spectrum of $\tw$ by a dashed
histogram in \fig{fig:harm}.
The tight agreement between the solid and dotted histograms in small
wave number strongly support our claim.

\section{Discussion}
\label{sec:discussion}

In this paper, we use a three-scale model to derive the saturation
time scales and saturation levels of the condensate and turbulent
fluctuation in 2D hydrodynamic turbulence.
This requires integrating the 2D Navier--Stokes equations for a very
long time.
This has been made possible by virtue of the high performance of the
GPUs.
We use the Ornstein-Uhlenbeck process as a phenomenological model to
describe the movement of condensate vortices.
In terms of the saturation time scales $\tE$ and $\tZ$, the DNS agree
quite well with the analytical predictions.
The saturation levels are, however, off by a constant factor $\zeta =
0.76$.
We speculate that this disagreement is because our three-scale model
ignores the non-trivial shape of the coherent vortices.
The higher harmonics of the condensate modify the spectra at small
wave number and affect the saturation level.
We have checked that it is not possible to remedy this problem by just
changing $k_1$ in our model to an effective wave number
$k_\mathrm{cond} \gtrsim \c$.
A more sophisticated model is needed.

Before we conclude, we should point out that the direct (forward)
enstrophy flux,
\begin{align}
  \dZ \equiv \eta_\infty \approx \f^2 (\k^2 -\c^2),
\end{align}
is smaller than the enstrophy input $\f^2\k^2$ in our three-scale
model.
There is no inconsistency here.
Their difference is simply the (very small) inverse enstrophy
\emph{leakage}
\begin{align}
  \iZ \equiv \f^2\k^2 - \dZ \approx \f^2 \c^2
\end{align}
that we commented on for Eq.~\eq{eq:dZ'/dt}.
The property $\iZ \ll \dZ$ is nothing magical.
It is simply a consequence of the broken power law in our three-scale
model.

Similarly, we can check how the direct (forward) energy \emph{leakage}
$\dE$ compares with the inverse cascade $\iE$.
Recalling that $\e$ is only the inverse energy transfer rate at $\c$,
we need to add the contribution from $2\nu Z'$,
\begin{align}
  \iE &\equiv \e + 3\nu\gamma Z'_\infty \approx
  \f^2\left[1 - 
      \left(1 - \frac{3}{2}\gamma\right)\frac{\k^2}{\gamma\K^2}
      \right], \\
  \dE &\equiv 4\nu\gamma\ln\frac{\K}{\k} Z'_\infty \approx
  \f^2\left(1 - \frac{3}{2}\gamma\right)\frac{\k^2}{\gamma\K^2}.
\end{align}
Using the original definition~\eq{app:gamma}, it is easy to verify
that $1 - 3\gamma/2 \lesssim 1$, which allows us to conclude $\iE \gg
\dE$.
Most of the input energy is inversely cascaded and dissipated at the
condensate.
Note that the above approximations are made at the same order.
The fact that $\iE + \dE = \f^2$ holds, therefore, shows the
consistence of our model.

Finally, our work stresses the fact that, as in three-dimensional
turbulence~\citep{Turbulence}, the order of taking the limit of long
time ($t \to \infty$) and small viscosity ($\nu \to 0$) is also
crucial in 2D turbulence.
Specifically, taking the limit $\nu \rightarrow 0$ first leads to a
(linear) divergence in time
\begin{align}
  \lim_{t\rightarrow\infty}\left(\lim_{\nu\rightarrow 0} \E\right) =
  \lim_{t\rightarrow\infty}\f^2 t.
\end{align}
We must take the limits in the correct order, which is,
\begin{align}
  \lim_{\nu\rightarrow 0}\left(\lim_{t\rightarrow\infty} \E\right) &=
  \lim_{\nu\rightarrow 0}\left(\frac{\f^2}{2\nu\c^2}
  \frac{\gamma\K^2 - \k^2}{\gamma\K^2 - \c^2}\right) \nonumber\\
  & = \lim_{\nu\rightarrow 0}\;\frac{\f^2}{2\nu\c^2}.
\end{align}
The system can then reach a steady state as long as the viscosity
remains non-zero, even without the Ekman term.

\acknowledgments

All simulations presented in this paper were done on the \texttt{Zorn}
cluster in PDC and the \texttt{Platon} cluster in Lunarc.
CKC is supported by a NORDITA fellowship.
CKC and AB thank John Bowman and Malcolm Roberts for insightful
discussions, which motivated this work.
All three authors thank Alex Hubbard and two anonymous referees who
helped improve the paper.
We also thank the organizers of the \emph{Nature of Turbulence}
workshop and are grateful to the hospitality by UCSB and KITP.
Financial support from European Research Council under the AstroDyn
Research Project 227952 and the Swedish Research Council under the
grant 2011-5423 are gratefully acknowledged.

\bibliography{ms,my}

\end{document}